\documentclass{emulateapj}
\def\lsim{\raise0.3ex\hbox{$<$}\kern-0.75em{\lower0.65ex\hbox{$\sim$}}}
\def\gsim{\raise0.3ex\hbox{$>$}\kern-0.75em{\lower0.65ex\hbox{$\sim$}}}

\begin{document}
\title{Substellar-Mass Condensations in Prestellar Cores}
\author{Fumitaka Nakamura\altaffilmark{1}, Shigehisa Takakuwa\altaffilmark{2}, 
Ryohei Kawabe\altaffilmark{1}
}
\altaffiltext{1}{National Astronomical Observatory, Mitaka, Tokyo 181-8588, 
Japan; fumitaka.nakamura@nao.ac.jp}
\altaffiltext{2}{Academia Sinica Institute of Astronomy and 
Astrophysics, P.O. Box 23-141, Taipei 106, Taiwan}

\begin{abstract}
We present combined Submillimeter-Array (SMA) + single-dish images 
of the (sub)millimeter 
dust continuum emission toward two prestellar cores SM1 and B2-N5 
in the nearest star cluster forming region, $\rho$ Ophiuchus.
Our combined images indicate that SM1 and B2-N5 consist of three
and four condensations, respectively, with masses of 
$10^{-2}-10^{-1}M_\odot$ and sizes of a few hundred AU.
The individual condensations have mean densities of $10^8-10^9$ cm$^{-3}$
and the masses are comparable to or larger than the critical 
Bonner-Ebert mass, indicating that the self-gravity plays an 
important role in the dynamical evolution of the condensations.
The coalescence timescale of these condensations is estimated to be
about $10^4$ yr, which is comparable to  
the local gravitational collapse timescale, suggesting that 
merging of the condensations, instead of accretion, 
plays an essential role in the star formation process.
These results challenge the standard theory of star formation, 
where a single, rather featureless prestellar core collapses to form
at most a couple of condensations, each of which potentially evolves 
into a protostar that is surrounded by a rotating disk where planets 
are created.
\end{abstract}

\section{Introduction}

According to the standard theory of star formation, stars form out of 
dense cores embedded in molecular clouds
\citep[e.g.,][]{shu87,mckee07}. 
The dense cores prior to the protostellar formation, or prestellar
cores, are thought to be smooth with no condensations inside 
\citep{bacmann00}.
The gravitational collapse of such a core impedes the growth of 
the local density fluctuations, resulting in the formation of at most a 
couple of condensations, each of which potentially evolves into 
a protostar that is surrounded by a rotating disk where planets are created.
Recent millimeter and submillimeter observations 
have revealed that the core mass functions (CMFs) resemble 
the stellar initial mass function (IMF) 
\citep{motte98,stanke06,maruta10}.
The resemblance between the CMFs and IMF appears to claim  
that these cores are the direct precursors of individual 
stars or small stellar systems like binaries.

The above scenario of star formation implicitly assumes that the core 
has a smooth density distribution and does not contain significant 
condensations. 
This assumption appears to be supported by recent observations
by \citet{schnee10,schnee12}, who suggested that the prestellar cores
do not contain significant substructures on the basis of 
the interferometric observations with arcsecond resolution 
\citep[see also][]{olmi05}. 
\citet{maury10} studied substructures around several Class 0 sources 
with subarcsecond resolution and found no clear signs of 
fragmentation at 100 AU. 
On the other hand, several previous observations 
revealed the presence of gravitationally-unbound, 
substellar-mass structures with sizes of a few thousands AU 
inside prestellar cores in Taurus
\citep{langer95, peng98, takakuwa03, roy11}. 
The origins and fates of these small structures inside the prestellar
cores remain to be elucidated.
These previous studies of substructures in prestellar cores have
concentrated primarily on the relatively-isolated environments.
Hence, none of the previous observations have clearly answered 
the question as to whether prestellar cores in cluster-forming 
regions have significant substructure or not, 
because of the lack of high spatial resolution observations
[see, however, \citet{kamazaki01} for the discovery of 
thousand-AU scale, substellar-mass structures in Oph A]. 
However, since most stars form in the clustered environments, 
it is important to elucidate the structures and properties of prestellar cores 
in the clustered environments.

In the clustered environments, the identified cores tend 
to be more compact than those in quiescent regions, 
and their sizes are comparable to a telescope beam size.
An area covered by a typical single prestellar core 
in quiescent regions is often occupied by 
a small cluster of cores in cluster-forming regions \citep{motte98}. 
To verify whether prestellar cores in the cluster environments 
contain substantial internal structures or not,
we analyze in this Letter two representative prestellar cores 
(SM1 and B2-N5) in $\rho$ Ophiuchus, 
the nearest cluster-forming region at a distance of 120 pc,
using the dust continuum data obtained by SMA.
The SMA data have spatial resolution of a few hundreds AU, 
one of the finest among the available data of prestellar cores.

The first target SM1, located in the middle of the 
Oph A filamentary cloud, has the strongest submillimeter continuum emission
in $\rho$ Oph \citep{andre93}.
The filamentary morphology is suggestive of interaction
with a nearby young B star S1.
The total mass and dust temperature of SM1 are estimated to be 
2 $M_\odot$ and $T_{\rm d}\approx 20$ K, 
respectively \citep{motte98}.
The second target B2-N5 is located in the Oph B2 cloud 
that contains a number of prestellar cores.
In the Oph B2 cloud, a gigantic outflow has recently been 
detected \citep{nakamura11a}.  
This core was identified in the N$_2$H$^+$ (1-0) emission 
and is one of the cores having the largest column densities 
in the Oph B2 cloud.
The total mass and temperature are estimated to be about a few 
tenth $M_\odot$ \citep[N$_2$H$^+$]{friesen10} and $T\approx 15$ K 
\citep[NH$_3$]{friesen09}, respectively.

The rest of the paper is organized as follows.
The details of our data are described in Section \ref{sec:data}.
The results and discussion are presented in Sections 
\ref{sec:results} and \ref{sec:discussion}, respectively.

\section{Data}
\label{sec:data}

\subsection{Single-dish Continuum Data}

The $\rho$ Oph cluster-forming region consists of 
several sub-parsec scale clouds with masses of order 
of $10^2 M_\odot$, each of which contains several prestellar cores.
Our target cores SM1 and B2-N5 are located in 
the Oph A and B2 clouds, respectively. 
Here, we briefly describe the dust continuum data toward 
$\rho$ Oph, taken by the single-dish observations.

The 850 $\mu$m data of the Oph A cloud taken with the SCUBA 
were obtained through the archival system of the COMPLETE Survey
 \citep{ridge06b}. 
The FWHM beam size of the JCMT was 14$''$ at 850$\mu$m.
The mean rms noise level of the data is $\sim$10 mJy beam$^{-1}$.

The 1.1 mm data of the Oph B2 cloud were 
obtained with the AzTEC camera \citep{wilson08b} 
on the ASTE telescope \citep{ezawa04b}. 
The FWHM beam size of the ASTE was 28$''$ at 1.1 mm.
The observations were performed in the raster scan mode.
Each field was observed several times with azimuth and elevation scans. 
The separation among scans was adopted to be 117$''$, which is a
quarter of the AzTEC field of view (FoV $\sim$ 7$'.8$).
The effective beam size of the data was 40$''$ after the FRUIT imaging 
which is an iterative mapping method to recover the spatially-extended 
component \citep{liu10b}. 
The noise level was less than 10 mJy beam$^{-1}$ 
in the entire observed area. The details of the data will be
presented elsewhere.

In general, it is difficult to completely recover the emission 
from extended structures with bolometric observations using 
the ground-based telescopes due to the atmospheric emission. 
For SCUBA, the emission from the structures larger than $\sim 2'$ 
is suppressed in removing the atmospheric emission during the 
data reduction process, making the 850 $\mu$m map mostly 
devoid of the extended emission.
The AzTEC image also has the same problem for 
the structures larger than the field of view of $\sim 8'$. 
However, this effect is expected not to be severe
because the sizes of the cores and the condensations identified below are 
smaller than $\sim 2'$ and $8'$ for SCUBA and AzTEC, respectively.

\subsection{SMA Data}

The target prestellar cores SM1 and B2-N5 were observed 
in the SMA compact configuration at 870 $\mu$m and 1.1 mm, 
respectively. We obtained the continuum data through the SMA data archive.
The SMA is a double-sideband instrument \citep{ho04}, 
having a number of spectral windows of the SMA correlator (``chunks'').
Each target was observed in a single pointing mode.

Table \ref{tab:sma} summarizes the observational parameters.
The minimum projected baseline lengths were 11 and 12 $k\lambda$
(8.2$''$ and 7.5$''$) at 870 $\mu$m and 1.1 mm, respectively, and 
for a Gaussian emission distribution with a FWHM of $\sim 7''$ 
($\sim $ 800 AU), the peak flux recovered is $\sim$ 50\% of the peak flux 
of the Gaussian \citep{wilner94}.
This problem is called 
the missing flux problem. To recover the missing fluxes from extended
structure and obtain more reliable images and physical properties of the 
structures, we combined the SMA data and the available single-dish 
data \citep{takakuwa07}: i.e., 850 $\mu$m SCUBA
data for SM1 and 
1.1 mm AzTEC data for B2-N5.
Below, we mainly discuss the combined SMA$+$single-dish images.

\section{Results}
\label{sec:results}

\begin{figure}
\plotone{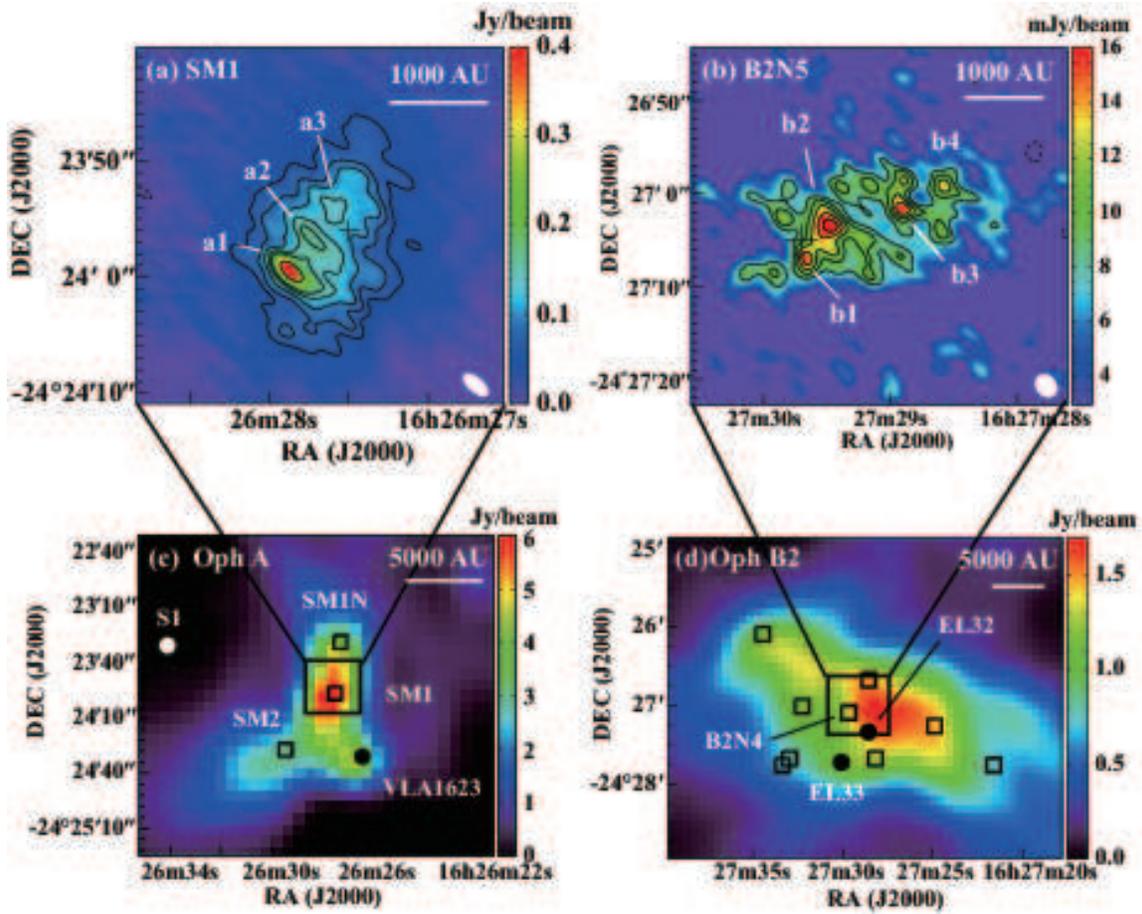} 
\caption{
(a) Combined SCUBA and SMA image toward SM1.
The contour levels are -20, 60, 80, 100, 130, 160, and 320 mJy/beam.
The cross is the position of SM1 identified by \citet{motte98}.
The alphabets designate the identified condensations.
(b) Combined AzTEC/ASTE and SMA image toward B2-N5.
The contour levels are -3, 8, 10, 12, 14, and 16 mJy/beam.
The cross is the position of B2-N5 identified by \citet{friesen10}.
(c) the SCUBA 850$\mu$m continuum image toward the Oph A 
region \citep{johnstone00}. 
The white and black filled circles are the B star, S1, and 
the prototypical Class 0 YSO, VLA 1623, respectively.
The positions of some submillimeter continuum sources are 
indicated by the squares \citep{motte98}.
(d) the AzTEC/ASTE 1.1 mm continuum image toward the Oph B2 region. 
The black filled circles are the positions of Class I YSOs, EL32 and EL33.
In panels (a) and (b), the synthesized beams are 
shown in the lower right of the panels.
}
\label{fig:sma+single}
\end{figure}

\begin{figure}
\plotone{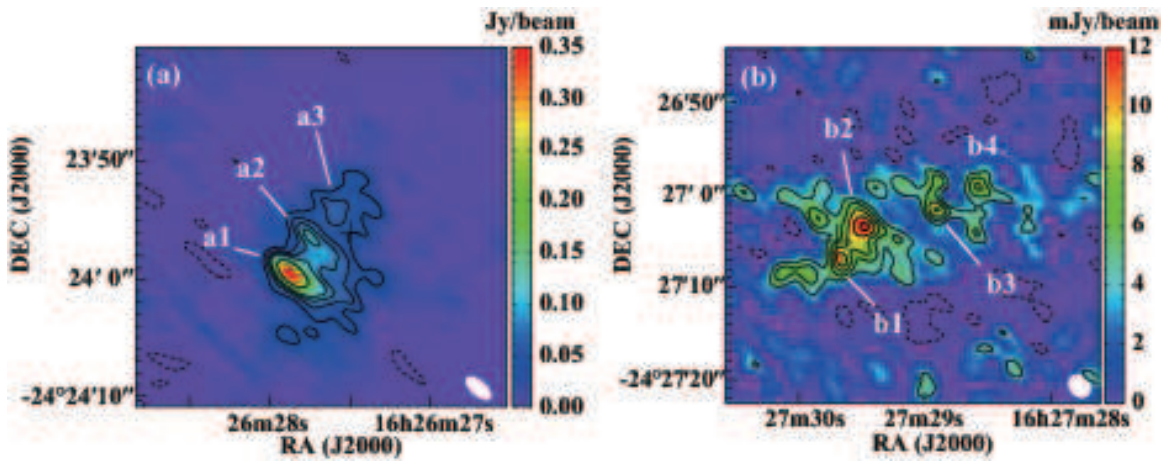} 
\caption{(a) 
Original SMA image toward SM1.
The contour levels are -20, 40, 60, 80, 100, and 200 mJy/beam.
The cross is the position of SM1.
The alphabets designate the condensations identified 
in Fig. \ref{fig:sma+single}(a).
(b) Original SMA image toward B2-N5.
The contour levels are -3, 4, 6, 8, 10, 12 mJy/beam.
The alphabets designate the condensations identified 
in Fig. \ref{fig:sma+single}(b).
The synthesized beam is shown in the lower right of each panel.
}
\label{fig:sma}
\end{figure}

The combined images are presented in Figure \ref{fig:sma+single}.
For comparison, the original SMA images are also shown in 
Fig. \ref{fig:sma}.
Our high spatial dynamic range images constructed by 
combining the single-dish data and SMA data have spatial resolution and
sensitivity to dust emission superior to those of existing
millimeter and submillimeter observations toward prestellar cores 
in active cluster-forming regions, without significant 
effects of the missing flux inherent in any interferometric observations.
Both images have clearly revealed the presence of significant 
substructure inside the prestellar cores. 
Here, we define a compact $10^2-10^3$ AU-scale structure
with a peak intensity larger than 10 $\sigma$ rms noise level
as a condensation.
For SM1, we identified three condensations, labeled 
as a1, a2, and a3 in Fig. \ref{fig:sma+single}(a).
The southernmost condensation, a1, appears extremely compact, 
and the internal structure is unresolved 
even with a few arcsecond resolution.
The existence of the thousands AU-scale substructure 
in the Oph A cloud has been pointed out by
the previous interferometric observations \citep{kamazaki01}.
Our image indicates that the substructure previously identified 
as the strongest 3mm continuum emission 
(core A in Figure 2 of \citet{kamazaki01}) is much more compact 
and consists of two condensations a1 and a2.
The substructure is more prominent in B2-N5, which contains several 
small condensations, as shown in Fig. \ref{fig:sma+single}(b).  
Here, we identified 4 condensations in the map.  
The separations of the condensations are about 5$-$10 arcseconds, 
or $10^3$ AU.

We note that the condensations identified from the combined images
can be recognized easily even in the interferometric
images alone (Figure \ref{fig:sma}), and the small-scale structures 
are essentially the same 
as those presented in Figure \ref{fig:sma+single}. 
The peak fluxes of all the condensations identified in the SMA images
alone are more than 4$\sigma$ rms noise levels, indicating that 
these condensations are the real structures existing in the prestellar cores.

Table \ref{tab:condensations} summarizes the physical properties 
of the identified condensations.
We applied two-dimensional Gaussian fitting to the condensations 
by using the IMFIT task in MIRIAD and derived their sizes, positions, 
and peak and total fluxes.
The masses of the condensations are estimated 
using the dust opacity of 
$\kappa_\lambda=0.1 (250\mu m/\lambda)^\beta$ g$^{-1}$ cm$^{2}$ with
$\beta=2$ \citep{hildebrand83}.
This opacity is in agreement with the values adopted by
other studies of prestellar cores within a factor of a few
\citep[e.g.,][]{ossenkopf94,evans09,andre96}.
The typical masses of the condensations are of the order of 
$10^{-2} - 10^{-1} M_\odot$.  
For comparison, the mass of the critical Bonner-Ebert sphere whose
mean density coincides with the estimated mean density 
of each condensation is listed in Table \ref{tab:condensations}. 
The elongated beam shapes influence the estimation of the local
densities and therefore $M_{\rm BE}$. Since the identified condensations 
have sizes comparable to the beam sizes, $M_{\rm BE}$ presented 
in Table \ref{tab:condensations} may be underestimated.
Almost all the condensations have masses comparable to or 
larger than the critical Bonner-Ebert masses ($M_{\rm BE}$), 
indicating that the self-gravity plays an important role 
in the dynamical evolution of the condensations.

We note that the masses estimated from the continuum emission 
depend strongly on the adopted opacity and temperature.
These two parameters are generally difficult to measure accurately.
The dust opacity has uncertainty at least by a factor of a few.
Therefore, the masses of the condensations listed in Table 
\ref{tab:condensations} are reduced by a factor of a few when
the actual dust opacity is smaller. 
The adopted temperatures also have uncertainty.
Here, $T = 20$ K is adopted for SM1, although a larger temperature 
of 27 K is reported \citep{andre93}. 
For B2-N5, the dust temperature is assumed to be equal to the kinetic
temperature of 15 K, which is determined by the NH$_3$ observations
\citep{friesen09}. 
In Oph B2, the protostellar outflow from EL32 
\citep{nakamura11a} may have directly injected
turbulent motions in adjacent dense regions including B2-N5,
and may have increased the temperature of B2-N5.
However, the N$_2$H$^+$ ($J=1-0$) hyperfine fitting 
indicates that the temperature of B2-N5 appears low \citep{friesen09}.
The NH$_3$ observations may not trace the densest gas in Oph B2, 
and the adopted temperature may be overestimated in the densest parts.
Even if we take into account these uncertainties of the dust opacity and 
temperature, all the condensations still have masses that are 
comparable to or larger than $M_{\rm BE}$, for SM1.
For Oph B2, the masses of the condensations 
may become somewhat smaller than $M_{\rm BE}$ when we take into account 
the larger dust opacity and smaller temperature.
Even in that case, the existence of the large-amplitude fluctuations 
at this small-scale is surprising because
such large-amplitude fluctuations give a significant impact on 
the structure formation during the gravitational collapse of 
the parent core \citep{goodwin04}.

\section{Discussion}
\label{sec:discussion}

The significant substructure inside the  prestellar
cores implies that 
the basic unit of star formation in the clustered environments is 
much smaller than previously considered. 
In fact, recent theoretical and observational
studies suggest that prestellar cores formed in the clustered 
environments tend not to be self-gravitating, but pressure-confined
because of the strong inter-core 
turbulence \citep{maruta10,nakamura11b}.

\begin{figure}
\epsscale{0.7}
\plotone{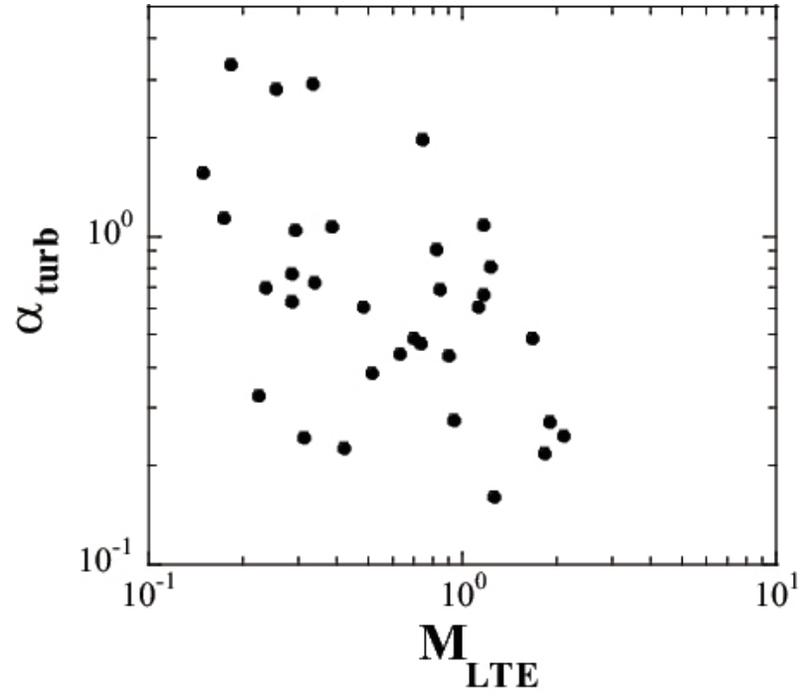} 
\caption{
The ratio of turbulent to gravitational energy ($\alpha_{\rm turb}$)
as a function of LTE mass ($M_{\rm LTE}$)
for the $\rho$ Oph prestellar cores identified from H$^{13}$CO$^+$ (1-0)
 \citep{maruta10}.
}
\label{fig:alpha}
\end{figure}

How were these substellar-mass condensations created in parent 
prestellar cores? One possibility is the fragmentation due to 
internal turbulent motions, or turbulent fragmentation, in short.  
In the standard theory of star formation, the cloud turbulence 
is considered to be less important inside the prestellar cores.
However, cluster-forming regions are highly turbulent.
Supersonic turbulence potentially triggers the formation of dense 
regions where the turbulent dissipation preferentially 
happens \citep{nakamura11b}.  
In such dense regions the large-amplitude fluctuations can be left.
To quantify the dynamical importance of the core internal turbulence,
we present in Figure \ref{fig:alpha}
the ratios of the turbulent to gravitational energy 
($\alpha_{\rm turb} = U / W$) for the prestellar cores 
in $\rho$ Oph as a function of the LTE mass, $M_{\rm LTE}$, 
where $U$ is the turbulent energy $U = M_{\rm LTE} \delta v_{\rm NT}^2/2$, 
$W$ is the gravitational energy $W=GM_{\rm LTE}^2/R_{\rm core}$, 
 $\delta v_{\rm NT}$ is the three-dimensional nonthermal velocity dispersion, 
$G$ is the gravitational constant, and $R_{\rm core}$ is the core
radius.  
All the prestellar cores were identified in a high density tracer, 
H$^{13}$CO$^+$ ($J=1-0$), using the Nobeyama 45m 
telescope whose beam size is about 2000 AU \citep{maruta10}.
The mean value of $\alpha_{\rm turb}$ is estimated to be about 0.8, 
suggesting that the internal turbulence can contribute significantly 
to the core dynamics.  
This is in contrast to the cores of quiescent regions 
like Taurus \citep{goodwin04}, where $\alpha_{\rm turb}$ tends to be
smaller.
In addition, the values of $\alpha_{\rm turb}$ are significantly 
larger than the threshold for turbulent fragmentation of 
$\alpha_{\rm turb} \gsim 0.03$ \citep{goodwin04}.
Thus, the turbulent motions are responsible for the formation of 
the condensations inside the prestellar cores in $\rho$ Oph.

H$^{13}$CO$^+$ sometimes suffers from the molecular depletion 
in the dense parts of the prestellar cores. 
Since \citet{maruta10} adopted a constant H$^{13}$CO$^+$ 
fractional abundance, the estimated core masses (and thus $\alpha_{\rm turb}$) 
may be underestimated (overestimated)
if H$^{13}$CO$^+$ is significantly depleted.
However, the depletion effect is expected to be minor 
as discussed in \S 4.3 of \citet{maruta10}.
Therefore, we believe that the estimation of the core masses 
and $\alpha_{\rm turb}$ is reasonable, and our conclusion is not 
altered by the effects of the molecular depletion.

What is the fate of these substellar-mass condensations?  
Both the prestellar cores  contain condensations
with sizes of order $10^2$ AU and masses of a few hundredth $M_\odot$. 
Almost all the condensations have masses comparable 
to or larger than the critical Bonner-Ebert mass. 
Therefore, they are expected to be gravitationally bound.
From the mean separation between the condensations of  
a few $\times 10^2$ AU
the coalescence timescale of these condensations is estimated 
to about $10^4$ yr, comparable to the free-fall times 
of the individual condensations with densities of $10^{8}$ cm$^{-3}$, 
where the relative velocities among condensations are assumed to be
equal to the velocity dispersion of B2-N5, $\delta v\sim 0.2$ km s$^{-1}$.  
Therefore, these condensations can either merge with themselves
or collapse individually to form small groups of 
low-mass protostars and/or prot-brown dwarfs. 
Here, the mean separation projected on the plane of the sky
is adopted to calculate the coalescence timescale and therefore the 
estimated coalescence timescale gives a lower limit.
Unless the actual mean separation is larger than the adopted value
by a factor of $\sim 10$, the coalescence timescale is still 
comparable to the local free-fall time and thus 
our conclusion is not altered significantly
(see \citet{peng98} and \citet{takakuwa03} for more quantitative 
analysis on the separations of condensations.).

According to the standard theory of star formation, 
a typical prestellar core has sizes of 10$^4$ AU and 
masses of a few $M_\odot$ $-$ 10 $M_\odot$ 
and is considered to have no significant substructures.
The fragmentation during the prestellar phase 
is considered to be significantly impeded by the global gravitational 
collapse before 
a opaque disk is formed \citep{hennebelle08,machida08}.  
The prestellar cores could fragment into at most 
a couple of condensations that grow in mass by accretion and eventually 
evolve into protostars. Thus, a binary or small multiple system
is created.  The rotating disks are created around individual
protostars and the planets are produced there.
However, the recent hydrodynamic simulation of gravitational 
collapse of a prestellar core has posed a challenge to this standard
picture.  In a series of three-dimensional radiation hydrodynamic 
simulations, the protostellar radiation destroys 
the opaque disks that would evolve into circumstellar 
disks \citep{schonke11}. 
Although whether the disks are completely destroyed or not may 
depends on the size of the disk formed, 
this effect 
of the protostellar radiation significantly constrains
the planet formation from the circumstellar disks.
In our scenario, substellar-mass condensations are already 
formed before the protostars are created.  
Some of these substellar-mass condensations potentially 
evolve into giant planets 
orbiting around protostars or proto-brown dwarfs under the 
influence of the protostellar radiation that can remove a significant
amount of mass from the condensations.

Here, we revealed that the two representative prestellar cores
in the $\rho$ Oph cluster forming region contain 
hundreds AU-scale structures. 
Another compact structure with $\sim$1000 AU was 
also found toward the southern part of Oph A, Oph A-N6
\citep{bourke12}.
Very recently, \citet{barsony12} discovered a 
surprising number of brown dwarf candidates in this region.
Although the follow-up spectroscopic observations would be 
needed, they estimated the ratio of the brown dwarfs 
($M \lesssim 0.1 M_\odot$) to low-mass stars 
($0.1 M_\odot \lesssim M \lesssim 1 M_\odot$) 
to be about 1.4, more than 5 times larger than the previous estimation.
Our observations may explain the abundance of substellar-mass 
objects like brown dwarfs in this region.
However, it remains unclear whether the small-scale structures 
are really common in prestellar cores in the clustered environments.
Further observations with higher spatial resolution and higher
sensitivity will be needed to clarify the internal structures 
of the prestellar cores. 
The high-angular resolution interferometric observations such as ALMA 
will provide us an excellent opportunity to explore the substructures of 
prestellar cores.

\acknowledgements
We thank the referee for valuable comments that improve the paper.
This work was supported partly by a Grant-in-Aid for 
Scientific Research of Japan, 20540228 and 22340040 (F. N.)
and a grant from the National Science Council of 
Taiwan NSC 99-2112-M-001-013-MY3 (S. T.).



\begin{deluxetable}{lll}
\tabletypesize{\scriptsize}
\tablecolumns{3}
\tablecaption{Parameters for the SMA Observations of SM1 and B2-N5}
\tablewidth{0.8\columnwidth}
\tablehead{\colhead{Parameter} 
& \colhead{SM1} & \colhead{B2-N5} 
}
\startdata
Observing Date & July 29, 2007 & June 1, 2008 \\
Number of antennas & 7 & 8 \\
R.A. (J2000) & 16:26:27.60 & 16:27:29.15 \\
Decl.(J2000) & -24:23:55.0 & -24:27:02.0 \\
Primary beam HPBW & 36$''$ & 45$''$ \\
Synthesized beam HPBW &$2.7'' \times 1.3''$ (P.A.=49$^\circ$) 
   & $2.7'' \times 2.2''$ (P.A.=40$^\circ$) \\
Baseline coverage & 10.9 - 148.2 (k$\lambda$) & 12.1 - 114.2 (k$\lambda$) \\
LO frequency & 341.5 GHz & 274.2 GHz \\
Bandwidth & 4.0 GHz  & 3.8 GHz \\
Gain calibrator & NRAO530 & 1625-254  \\
Flux of the gain calibrator & 1.4 Jy & 0.79 Jy \\
Passband calibrator & 3C273, 3C454.3 & 3C279, Jupiter \\
Flux calibrator & Uranus &  Titan \\
System temperature & 150 - 600 (K) & 300 - 600 (K)  \\
rms noise level & 4.8 mJy/beam & 1.2 mJy/beam 
\enddata
\label{tab:sma}
\end{deluxetable}

\begin{deluxetable}{lllllllll}
\tabletypesize{\scriptsize}
\tablecolumns{9}
\tablecaption{Physical properties of condensations in prestellar cores.
}
\tablewidth{\columnwidth}
\tablehead{\colhead{Name} 
& \colhead{R.A.} & \colhead{Dec.} & \colhead{S$_{\rm p}$} 
& \colhead{S$_{\rm t}$} 
& \colhead{Size} & \colhead{Mass} & \colhead{density} 
& \colhead{$M_{\rm BE}$} \\
\colhead{} & \colhead{(J2000)} & \colhead{(J2000)} & \colhead{(mJy/beam)} 
& \colhead{(mJy)} 
& \colhead{(AU$\times$AU)} & \colhead{($M_\odot$)} 
& \colhead{($\times 10^8$ cm$^{-3}$)} 
& \colhead{ ($M_\odot$)} 
}
\startdata
a1 & 16:26:27.8 & -24:24:00 & 455 & 742 & $350\times 167$ &0.15  & 30.7
			     & 0.014 \\
a2 & 16:26:27.7 & -24:23:57 & 176 & 958 & $770\times 287$ & 0.17  & 4.6
			     & 0.036 \\
a3 & 16:26:26.6 & -24:23:54 & 138 & 769 & $591\times 532$ & 0.13 & 2.2
			     & 0.053\\
\hline
b1 & 16:27:29.6 & -24:27:07 & 15.2 & 44 & $532\times 310$ & 0.033 & 1.4
			     & 0.03 \\
b2 & 16:27:29.5 & -24:27:04 & 17.4 & 136 & $960\times 702$ &0.10 &
			     0.53 & 0.056 \\
b3 & 16:27:28.9 & -24:27:02 & 15.8 & 47 & $548\times 340$ & 0.035 &
    1.3 & 0.036 \\
b4 & 16:27:28.6 & -24:26:59 & 13.1 & 49 & $541\times 480$ & 0.036 & 0.79
& 0.046 
\enddata
\label{tab:condensations}
\tablecomments{The physical quantities listed in the table are evaluated 
from deconvolved Gaussian-fitted model parameters. 
The core masses are estimated by adopting the dust temperatures 
of 20 K and 15 K for the prestellar cores, SM1 and B2-N5, respectively.
The size is given by the FWHM major and minor axes obtained
from the IMFIT Gaussian fitting.}
\end{deluxetable}

\end{document}